\documentclass[12pt,a4paper]{article}
\usepackage{graphicx}
\usepackage{subeqnarray}
\begin{document}
%
%
%
%
\def\astrobj#1{#1}
\newenvironment{lefteqnarray}{\arraycolsep=0pt\begin{eqnarray}}
{\end{eqnarray}\protect\aftergroup\ignorespaces}
\newenvironment{lefteqnarray*}{\arraycolsep=0pt\begin{eqnarray*}}
{\end{eqnarray*}\protect\aftergroup\ignorespaces}
\newenvironment{leftsubeqnarray}{\arraycolsep=0pt\begin{subeqnarray}}
{\end{subeqnarray}\protect\aftergroup\ignorespaces}
\newcommand{\diff}{{\rm\,d}}
\newcommand{\pprime}{{\prime\prime}}
\newcommand{\sgn}{{\rm sgn}}
\newcommand{\szeta}{\mskip 3mu /\mskip-10mu \zeta}
\newcommand{\stau}{\mskip 3mu /\mskip-10mu \tau}
\newcommand{\semme}{\mskip 3mu /\mskip-12mu m}
\newcommand{\FC}{\mskip 0mu {\rm F}\mskip-10mu{\rm C}}
\newcommand{\appleq}{\stackrel{<}{\sim}}
\newcommand{\appgeq}{\stackrel{>}{\sim}}
\newcommand{\legr}{\stackrel{<}{>}}
\newcommand{\grle}{\stackrel{>}{<}}
\newcommand{\Int}{\mathop{\rm Int}\nolimits}
\newcommand{\Nint}{\mathop{\rm Nint}\nolimits}
\newcommand{\range}{{\rm -}}
\newcommand{\displayfrac}[2]{\frac{\displaystyle #1}{\displaystyle #2}}
%
%
\title{
A double power-law fit to the computed stellar
$\log(\tau/{\rm y})$-$\log(m/m_\odot)$ relation
}
\author{{$\qquad~$R.~Caimmi}\footnote{
{\it Physics and Astronomy Department, Padua Univ., Vicolo Osservatorio 3/2,
I-35122 Padova, Italy}
email: roberto.caimmi@unipd.it~~~
fax: 39-049-8278212}
\phantom{agga}}
%
%
\maketitle
\begin{quotation}
\section*{}
\begin{Large}
\begin{center}
Abstract

\end{center}
\end{Large}
\begin{small}

\noindent\noindent

The computed $\log(\tau/{\rm y})$-$\log(m/m_\odot)$ relation for the stellar
initial mass range, $0.6\le m/m_\odot\le120.0$, and the stellar initial
metallicity range, $0.0004\le Z\le0.0500$, tabulated in an earlier attempt
(Portinari et al. 1998) is fitted to a good extent by a four-parameter curve,
expressed by a double power-law, for assigned stellar initial metallicity,
which can be reduced to a three-parameter curve, expressed by a single
power-law, for
the whole set of stellar initial metallicities.   The relative errors,
$R[\log(\tau/{\rm y})]=1-\log(\tau_{\rm fit}/{\rm y})/\log(\tau/{\rm y})$, do
not exceed about 2\% and 4\%, respectively.   The extent to which the
interpolation curve, expressed by a single power-law, can be extrapolated
towards both high-mass and low-mass stars, is also investigated.   High-mass
star lifetimes are understimated by a factor less than 2 up to $m/m_\odot=
1000$ and by a fiducial factor less than 4 up to $m/m_\odot\to+\infty$.
Low-mass star lifetimes are overstimated by a factor of about 3 down to
$m/m_\odot=0.25$ and by an unacceptably large factor down to $m/m_\odot=0.08$.
As a simple application, the star mass fraction of a single
star generation with stellar initial mass function defined by a power-law, is
plotted vs. the logarithmic stellar lifetime.   The star mass
fraction declines in time at a decreasing rate for mild stellar initial mass
function and at an increasing rate for steep stellar initial
mass function, where a linear trend is exhibited for a value
of the exponent close to the Salpeter's value $(p_{\rm lin}\approx-2.35)$.

\noindent
{\it keywords -
stars: evolution - stars: formation.}
\end{small}
\end{quotation}
\pagebreak

\section{Introduction} \label{intro}

For isolated stars, the stellar lifetime is defined as the time needed to move
from the zero-age main sequence up to the giant branch and through any
subsequent giant evolution.   There is no easy way to infer stellar lifetime
as a function of stellar initial mass, from a selected star sample belonging
to an assigned population.   Accordingly, $\tau$-$m$ or
$\log(\tau/{\rm y})$-$\log(m/m_\odot)$ relations, where $\tau$ is the stellar
lifetime and $m$ the stellar initial mass, are interpolation curves to the
results from stellar evolution models (e.g., Maeder and Meynet 1989; Padovani
and Matteucci 1993).   For further details, an interested reader is addressed
to recent overviews (e.g., Romano et al. 2005).

A systematic dependence of the $\log(\tau/{\rm y})$-$\log(m/m_\odot)$ relation
on the stellar initial metallicity, $Z$, is available in tabular form within
the mass range, $0.6\le m/m_\odot\le120.0$, and the initial metallicity range,
$0.0004\le Z\le0.0500$, where the changes due to different initial
metallicities can be neglected to a first extent (Portinari et al. 1998).
Accordingly, lifetime is a strongly decreasing function of the initial mass
and only weakly dependent on the initial metallicity (e.g., Wiersma et al.
2009).

Interpolating the computed $\log(\tau/{\rm y})$-$\log(m/m_\odot)$ relation
via a simple, continuous, derivable curve, could be useful in problems where
the first derivatives, $\diff\tau/\diff m$ or $\diff m/\diff\tau$, and higher
order derivatives are needed, such as chemical evolution models.   The current
note focuses on a computed $\log(\tau/{\rm y})$-$\log(m/m_\odot)$ relation
where the dependence of stellar lifetime on stellar initial mass and
metallicity is exploited (Portinari et al. 1998), aiming to the expression of
four-parameter curves involving a double power-law, more
specifically an exponential whose argument is, in turn, a power.

To this respect, using standard regression techniques seems to be
inappropriate, in that
different stellar evolution models yield different results for an assigned
choice of input parameters, due to still persisting uncertainties on some
physical processes, and standard regression curve procedures would be
of little significance in the case under discussion.   Then a different
strategy must be exploited, where four ``crossing'' points are specified as
belonging to the interpolation curve.

With regard to a fixed stellar initial metallicity, the coordinates of the
crossing points can coincide with computer outputs,
$[\log(m_{\rm U}/m_\odot),\log(\tau_{\rm U}/{\rm y})]$, U = 0, A, B, C, to be
fixed
as appropriate.   With regard to the whole set of initial metallicities for
which the results are available, the ordinates of the crossing points can be
determined as average values of their counterparts related to the whole set of
initial metallicities, while abscissae remain unchanged,
$[\log(m_{\rm U}/m_\odot),\overline{\log(\tau_{\rm U}/{\rm y})}]$, U = 0, A,
B, C, to be fixed as appropriate, where $\overline{\log(\tau_{\rm U}/{\rm y})}
=[\log(\tau_{\rm U}/{\rm y})_1+...+\log(\tau_{\rm U}/{\rm y})_N]/N$ and $N=5$
in the case under discussion (Portinari et al. 1998).

Four-parameter interpolation curves, for both selected initial metallicities
and the whole set of initial metallicities, are derived in section \ref{fopa}.
The results are presented in section \ref{resu}.
An illustrative application is shown in section \ref{fras}.   The discussion
and the conclusion are performed in section \ref{dico}.

\section{Four-parameter interpolation curves} \label{fopa}

A complete set of computed stellar lifetime within the initial mass range,
$0.6\le m/m_\odot\le120.0$, and the initial metallicity range, $0.0004\le Z\le
0.0500$, is available in tabular form (Portinari et al. 1988).   More
specifically, stellar evolution has been computed for 30 stellar initial
masses, where 15 lie within the range, $0.6\le m/m_\odot\le2.0$, 7 within the
range, $2.5\le m/m_\odot\le9.0$, 6 within the range, $12.0\le m/m_\odot\le
60.0$, with the addition of $m/m_\odot=100.0, 120.0$; initial metallicities
have been fixed to $Z_1=0.0004$, $Z_2=0.0040$, $Z_3=0.0080$, $Z_4=0.0200$,
$Z_5=0.0500$.   It is worth noticing $Z_4$, initially conceived as solar
value, is supersolar according to a recent investigation yielding
$Z_\odot=0.134$ (Asplund et al. 2009).

\begin{figure}[t]  
\begin{center}      
\includegraphics[scale=0.8]{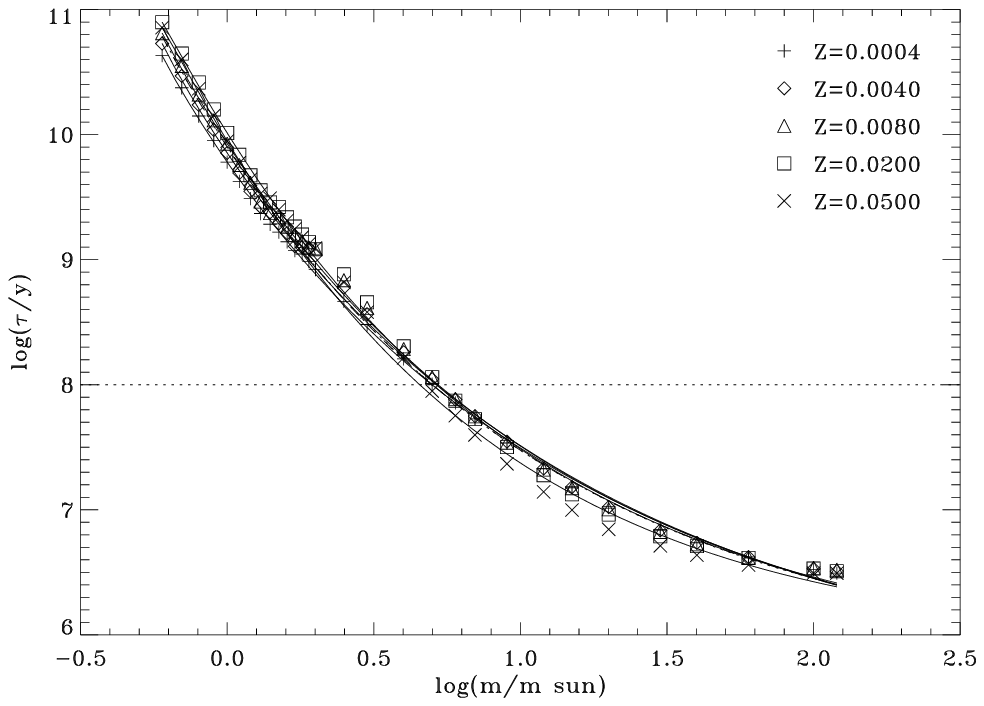}                      
\caption[ddbb]{Computed $\log(\tau/{\rm y})$-$\log(m/m_\odot)$ relation for 30
different stellar initial masses within the range, $0.6\le m/m_\odot\le120.0$,
in connection with 5 different stellar initial metallicities (Portinari et al.
1998).   Different symbols relate to different metallicities as indicated.
Four-parameter interpolation curves are shown as full for selected stellar
initial
metallicity; dashed for ordinates of crossing points averaged over the whole
set of stellar initial metallicity; dotted for parameters, $C_1$, $C_2$,
$C_3$, as in the last case and $\gamma$ set equal to unity.   The dotted
horizontal line marks a conventional boundary between long-lived (top) and
short-lived (bottom) stars.   See text for further details.}
\label{f:stelle}     
\end{center}       
\end{figure}

The computed $\log(\tau/{\rm y})$-$\log(m/m_\odot)$ relation, listed in
tabular form in an earlier attempt (Portinari et al. 1998), is plotted in
Fig.\,\ref{f:stelle} where different symbols relate to different initial
metallicities: $Z_1$ - crosses; $Z_2$ - diamonds; $Z_3$ - triangles; $Z_4$ -
squares; $Z_5$ - saltires.

The interpolation of the results shall be performed via four-parameter curves
defined as (Caimmi 2006):
\begin{lefteqnarray}
\label{eq:y}
&& y=f(x)=C_1\exp(-C_2x^\gamma)+C_3~~;\qquad x\ge0~~;
\end{lefteqnarray}
where $C_1$, $C_2$, $C_3$, $\gamma$, are parameters to be determined by
fitting to a good extent.

Regression curves based on standard methods, such as least squares
procedure, could be of little significance in that different stellar evolution
codes with same input parameters yield different results, due to still
persisting uncertainties on specific details of stellar evolution theory.

For this reason, a different strategy shall be adopted, namely interpolation
curves are constrained to pass through four crossing points, $(x_0,y_0)$,
$(x_A,y_A)$, $(x_B,y_B)$, $(x_C,y_C)$, which implies the determination of the
unknown parameters.   More specifically, the substitution of the coordinates
in Eq.\,(\ref{eq:y}) yields a system of four transcendental equations in four
unknowns, $C_1$, $C_2$, $C_3$, $\gamma$, as:
\begin{lefteqnarray}
\label{eq:y0}
&& y_0=C_1+C_3~~;\qquad x_0=0~~; \\
\label{eq:yU}
&& y_{\rm U}=C_1\exp(-C_2x_{\rm U}^\gamma)+C_3~~;\qquad {\rm U=A,B,C}~~;
\end{lefteqnarray}
where $x_0=0$, for convenience.

The combination of Eqs.\,(\ref{eq:y}) and (\ref{eq:y0}) yields:
\begin{lefteqnarray}
\label{eq:yy0}
&& \frac{y-C_3}{C_1}=\frac{C_1-y_0+y}{C_1}=[\exp(-C_2)]^{x^\gamma}>0~~;
\end{lefteqnarray}
which is equivalent to:
\begin{lefteqnarray}
\label{eq:C2}
&& C_2=x^{-\gamma}\ln\frac{C_1}{C_1-y_0+y}~~;
\end{lefteqnarray}
where $C_1<0$ or $C_1>y_0-y>0$ in the case under discussion.

The particularization of Eq.\,(\ref{eq:C2}) to $(x_U,y_U)$, $(x_V,y_V)$, after
performing the ratio on both sides of related equations, produces:
\begin{lefteqnarray}
\label{eq:xVU}
&& \left(\frac{x_{\rm V}}{x_{\rm U}}\right)^\gamma=\frac
{\ln[(C_1-y_0+y_{\rm V})/C_1]}{\ln[(C_1-y_0+y_{\rm U})/C_1]}~~;
\end{lefteqnarray}
with the above mentioned restrictions for $y_{\rm U}$, $y_{\rm V}$, to ensure
positive argument of related logarithm.   In the case under discussion, both
$x$ and $\gamma$ are real, which implies the ratio on the right-hand side of
Eq.\,(\ref{eq:xVU}) is non negative i.e. the arguments of the logarithms are
both either larger or lower than unity.   The related condition reads:
\begin{lefteqnarray}
\label{eq:yVU0}
&& \max(y_{\rm U},y_{\rm V})<y_0~~;\qquad\min(y_{\rm U},y_{\rm V})>y_0~~;
\end{lefteqnarray}
which holds in the case under discussion, keeping in mind
$y_{\rm U}<y_{\rm V}<y_0$ with no loss of generality, as shown in
Fig.\,\ref{f:stelle}.

Taking the logarithm on both sides of Eq.\,(\ref{eq:xVU}) yields:
\begin{lefteqnarray}
\label{eq:lnx}
&& \gamma\ln\frac{x_{\rm V}}{x_{\rm U}}=\ln\ln\frac{C_1-y_0+y_{\rm V}}{C_1}-
\ln\ln\frac{C_1-y_0+y_{\rm U}}{C_1}~~;
\end{lefteqnarray}
where U = A; V = B, C; with no loss of generality.   The cross product of
related equations reads:
\begin{lefteqnarray}
\label{eq:intc}
&& \ln\frac{x_{\rm C}}{x_{\rm A}}\left[\ln\ln\frac{C_1-y_0+y_{\rm B}}{C_1}-
\ln\ln\frac{C_1-y_0+y_{\rm A}}{C_1}\right] \nonumber \\
&& =\ln\frac{x_{\rm B}}{x_{\rm A}}\left[\ln\ln\frac{C_1-y_0+y_{\rm C}}{C_1}-
\ln\ln\frac{C_1-y_0+y_{\rm A}}{C_1}\right]~~;
\end{lefteqnarray}
which, in particular, can be solved for the crossing points, $(0,y_{\rm 0})$,
$(x_{\rm A},y_{\rm A})$, $(1,y_{\rm B})$, $(2,y_{\rm C})$, where
$x_{\rm B}=1$, $x_{\rm C}=2$, for convenience.

The intersection between the curve on the left and right-hand side of
Eq.\,(\ref{eq:intc}) yields the value of $C_1$.  The value of $C_3$ can be
readily inferred from Eq.\,(\ref{eq:y0}) as:
\begin{lefteqnarray}
\label{eq:C3}
&& C_3=y_0-C_1~~;
\end{lefteqnarray}
and the value of $C_2$ is determined via particularization of
Eq.\,(\ref{eq:yU}) to $(1,y_{\rm B})$.   The result is:
\begin{lefteqnarray}
\label{eq:C2p}
&& C_2=\ln\frac{C_1}{C_1-y_0+y_{\rm B}}~~;
\end{lefteqnarray}
finally, $\gamma$ is determined via particularization of Eq.\,(\ref{eq:lnx})
to U = A; V = B, C.   The result is:
\begin{lefteqnarray}
\label{eq:gamma}
&& \gamma=\left(\ln\frac{x_{\rm B}}{x_{\rm C}}\right)^{-1}\left[\ln\ln\frac
{C_1-y_0+y_{\rm C}}{C_1}-\ln\ln\frac{C_1-y_0+y_{\rm B}}{C_1}\right]~~;
\end{lefteqnarray}
where $x_{\rm B}/x_{\rm C}=1/2$ in the case under discussion.

With regard to interpolation curves related to selected stellar initial
metallicities, the expression of the mathematical variables, $(x,y)$, in terms
of the physical variables, $[\log(m/m_\odot),\log(\tau/{\rm y})]$, is
constrained by the domain of the interpolation curve, defined by
Eq.\,(\ref{eq:y}), $x\ge0$, which implies the following:
\begin{lefteqnarray}
\label{eq:xy}
&& x=\log\left(\frac m{m_\odot}\right)-\log\left(\frac{m_0}{m_\odot}\right)~~;
\qquad y=\log\left(\frac\tau{\rm y}\right)~~;
\end{lefteqnarray}
where $m_0$ is the sample star exhibiting the lowest mass i.e. the longest
lifetime, $m_0/m_\odot=0.6$ in the case under discussion (Portinari et al.
1998).   The crossing points can be explicitly expressed as:
${\sf P}_{\rm 0}\equiv[0,       \log(\tau_{0.6}/ {\rm y})]$,
${\sf P}_{\rm A}\equiv[-\log0.6,\log(\tau_{1.0}/ {\rm y})]$,
${\sf P}_{\rm B}\equiv[1,       \log(\tau_{6.0}/ {\rm y})]$,
${\sf P}_{\rm C}\equiv[2,       \log(\tau_{60.0}/{\rm y})]$,
where $\tau_{m/m_\odot}$ is the lifetime of a star with selected initial
metallicity and initial mass equal to $m$ in solar units.

With regard to interpolation curves related to ordinates of crossing points
averaged over the whole set of stellar initial metallicity, the expression of
the mathematical variables, $(x,y)$, in terms of the physical variables,
$[\log(m/m_\odot),$
$\overline{\log(\tau/{\rm y})}]$, for the abscissa, $x$, is
provided by Eq.\,(\ref{eq:xy}), while the ordinate, $y$, reads:
\begin{lefteqnarray}
\label{eq:yma}
&& y=\overline{\log\left(\frac\tau{\rm y}\right)}=\frac15\sum_{i=1}^5
\log\left(\frac\tau{\rm y}\right)_{Z_i}~~;
\end{lefteqnarray}
where $Z_i$, $1\le i\le5$, are the stellar initial metallicities for which
stellar evolution has been computed (Portinari et al. 1998).
The crossing points can be explicitly expressed as:
${\sf P}_{\rm 0}\equiv[0,       \overline{\log(\tau_{0.6}/ {\rm y})}]$,
${\sf P}_{\rm A}\equiv[-\log0.6,\overline{\log(\tau_{1.0}/ {\rm y})}]$,
${\sf P}_{\rm B}\equiv[1,       \overline{\log(\tau_{6.0}/ {\rm y})}]$,
${\sf P}_{\rm C}\equiv[2,       \overline{\log(\tau_{60.0}/{\rm y})}]$.

Within the stellar initial mass range under consideration,
$0.6\le m/m_\odot\le120.0$, the stellar lifetime range reads
$6<\log(\tau/{\rm y})<11$, which implies $y\gg0$ via Eqs.\,(\ref{eq:xy}) and
(\ref{eq:yma}).   Accordingly, use can be made of the relative errors:
\begin{lefteqnarray}
\label{eq:R}
&& R(y)=1-\frac{y_{\rm fit}}y~~;
\end{lefteqnarray}
where $y$ is inferred from computer output via Eq.\,(\ref{eq:xy}) or
(\ref{eq:yma}), and $y_{\rm fit}$ is the counterpart of $y$ on the related
interpolation curve.   The results are fitted to a good extent provided
relative errors maintain small enough, within a few percent say.

\section{Results} \label{resu}

With regard to selected stellar initial metallicities, the interpolated
$\log(\tau/{\rm y})$-$\log(m/m_\odot)$ relation is plotted in
Fig.\,\ref{f:stelle} as
full curves, which exhibit negligible difference with the exception of the
supersolar stellar initial metallicity, $Z=Z_5$, where slightly lower values
are shown.   A dotted horizontal line, $\log(\tau/{\rm y})=8$, marks a
conventional boundary between long-lived (up) and short-lived (down) stars,
which could be of interest for simple chemical evolution models (e.g., Pagel
and Patchett 1975; Caimmi 2011) and type Ia supernova progenitor models (e.g.,
Acharova et al. 2013).

The parameters of the interpolated $\log(\tau/{\rm y})$-$\log(m/m_\odot)$
relation, $C_1$, $C_2$, $C_3$, $\gamma$, are listed in Table \ref{t:pain}.
\begin{table}
\caption{Parameters (p) of the interpolated
$\log(m/m_\odot)$-$\log(\tau/{\rm y})$
relation, $C_1$, $C_2$, $C_3$, $\gamma$, for selected stellar initial
metallicities, $Z=Z_1$-$Z_5$ with regard to cases (c) 1-5, and for ordinates
of crossing points averaged over the whole set of stellar initial metallicity,
case 6, with the addition of the special choice, $\gamma=1$, leaving the
remaining parameters unchanged, case 7.   See text for further details.}
\label{t:pain}
\begin{center}
\begin{tabular}{llllll} \hline
\multicolumn{1}{l}{p:} &
\multicolumn{1}{c}{$Z$} &
\multicolumn{1}{c}{$C_1$} &
\multicolumn{1}{c}{$C_2$} &
\multicolumn{1}{c}{$C_3$} &
\multicolumn{1}{c}{$\gamma$} \\
c   & & & & & \\
\hline
1 & $Z_1$    & 5.10551 & 0.785151 & 5.52594 & 0.96950 \\
2 & $Z_2$    & 5.23694 & 0.784218 & 5.49141 & 0.96686 \\
3 & $Z_3$    & 5.20510 & 0.824914 & 5.60580 & 0.98915 \\
4 & $Z_4$    & 5.08428 & 0.904558 & 5.81444 & 1.03125 \\
5 & $Z_5$    & 4.92130 & 0.995420 & 5.93483 & 1.05302 \\
6 & $\bar Z$ & 5.07899 & 0.862584 & 5.70612 & 1.00322 \\
7 & $\bar Z$ & 5.07899 & 0.862584 & 5.70612 & 1       \\ 
\hline
\end{tabular}                     
\end{center}                      
\end{table}                       
The parameters lie within narrow ranges, namely $4.92<C_1<5.24$;
$0.78<C_2<1.00$; $5.49<C_3<5.71$; $0.96<\gamma<1.06$.   The special case,
$\gamma=1$, makes the interpolation curve reduce from a double power-law (an
exponential whose argument is, in turn, a power) to a simple exponential.
The interpolation curve related to averaged ordinates of crossing points and
its reduction to a simple exponential, cases 6 and 7 of Table \ref{t:pain},
are plotted in Fig.\,\ref{f:stelle} as a dashed and a dotted curve,
respectively.

\begin{figure}[t]  
\begin{center}      
\includegraphics[scale=0.8]{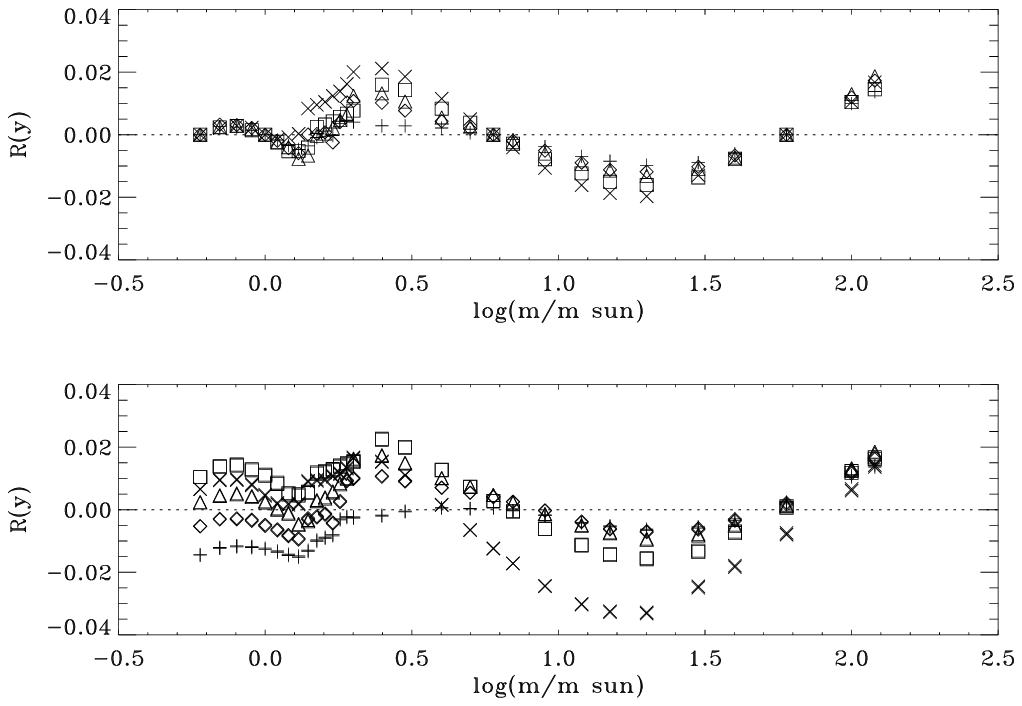}                      
\caption[ddbb]{The relative error, $R(y)=1-\log(\tau_{\rm fit}/{\rm y})/
\log(\tau/{\rm y})$, as a function of the logarithmic stellar initial mass,
$\log(m/m_\odot)$, with regard to interpolation curves related to selected
stellar initial metallicity (top panel) and to averaged ordinates of crossing
points (bottom panel).   Different symbols correspond to different stellar
initial
metallicity as shown in Fig.\,\ref{f:stelle}.   See text for further details.}
\label{f:stelrc}     
\end{center}       
\end{figure}

The relative errors, $R(y)$, defined by Eq.\,(\ref{eq:R}), as a function of
the logarithmic stellar initial mass, $\log(m/m_\odot)$, are plotted in
Fig.\,\ref{f:stelrc} for interpolation curves related to both selected
stellar initial metallicities, cases 1-5 of Table \ref{t:pain} (top panel) and
averaged ordinates of crossing points, cases 6-7 of Table \ref{t:pain} (bottom
panel), where different symbols are captioned as in Fig.\,\ref{f:stelle}.   An
inspection of Fig.\,\ref{f:stelrc} shows the relative errors do not exceed
about 2\% in the former alternative and less than 4\% in the latter one.
Then it may safely be thought interpolation curves, expressed by
Eq.\,(\ref{eq:y}), with values of parameters listed in Table \ref{t:pain},
satisfactorily fit to the computed $\log(\tau/{\rm y})$-$\log(m/m_\odot)$
relation.

The special case, $\gamma=1$, deserves further attention in that the
interpolation curve reduces to a simple exponential via Eq.\,(\ref{eq:y}), as:
\begin{lefteqnarray}
\label{eq:yexp}
&& y=C_1\exp(-C_2x)+C_3~~;
\end{lefteqnarray}
and the substitution of Eq.\,(\ref{eq:xy}) into (\ref{eq:yexp}) after some
algebra yields:
\begin{lefteqnarray}
\label{eq:logtau}
&& \log\left(\frac\tau{\rm y}\right)=C_1^\prime\left(\frac m{m_\odot}\right)^
{-C_2^\prime}+C_3~~; \\
\label{eq:Cp12}
&& C_1^\prime=C_1\exp\left[C_2\log\left(\frac{m_0}{m_\odot}\right)\right]~~;
\qquad\frac{m_0}{m_\odot}=0.6~~;\qquad C_2^\prime=\frac{C_2}{\ln10}~~;
\end{lefteqnarray}
which implies stellar lifetime is expressed by a double power-law, more
specifically an exponential whose argument is, in turn, a power, as:
\begin{lefteqnarray}
\label{eq:tame}
&& \frac\tau{\rm y}=\exp_{10}\left[C_1^\prime\left(\frac m{m_\odot}\right)^
{-C_2^\prime}+C_3\right]~~;
\end{lefteqnarray}
where the parameter values are:
\begin{lefteqnarray}
\label{eq:Cpv}
&& C_1^\prime=4.19439~~;\qquad C_2^\prime=0.374616~~;\qquad C_3=5.70612~~;
\end{lefteqnarray}
according to case 7 listed in Table \ref{t:pain}.

\section{A simple application} \label{fras}

With regard to a single star generation, let the stellar initial mass function
be a power-law as:
\begin{lefteqnarray}
\label{eq:IMF}
&& \phi\left(\frac m{m_\odot}\right)=A\left(\frac m{m_\odot}\right)^p~~;
\end{lefteqnarray}
where $A$ is the star formation efficiency and $-3\le p\le-2$ for the cases of
interest, in particular $p=-2.35$ relates to a classical investigation on the
solar neighbourhood (Salpeter 1955).

By definition,
$\diff N=\phi(m/m_\odot)\diff(m/m_\odot)$ is the number of stars born within
the mass range, $m/m_\odot\mp\diff(m/m_\odot)/2$, and the initial mass of the
star generation reads:
\begin{lefteqnarray}
\label{eq:M0}
&& M^\ast(0)=Am_\odot\int_{m_{\rm mf}/m_\odot}^{m_{\rm Mf}/m_\odot}\frac m
{m_\odot}\left(\frac m{m_\odot}\right)^p\diff\frac m{m_\odot}~~;
\end{lefteqnarray}
where $m_{\rm Mf}$ and $m_{\rm mf}$ are the upper and lower stellar initial
mass limit, respectively, in solar units.

The mass of the star generation (without stellar remnants) at the time,
$t=\tau(m/m_\odot)$, reads:
\begin{lefteqnarray}
\label{eq:Mt}
&& M^\ast\left(\frac\tau{\rm y}\right)=Am_\odot\int_{m_{\rm mf}/m_\odot}^
{m/m_\odot}\frac m{m_\odot}\left(\frac m{m_\odot}\right)^p\diff\frac m
{m_\odot}~~;
\end{lefteqnarray}
where $\tau(m/m_\odot)$ is the lifetime of a star of mass $m$ in solar units.

The fractional mass of the star generation,
$s(\tau/{\rm y})=M^\ast(\tau/{\rm y})/M^\ast(0)$, can be determined from 
Eqs.\,(\ref{eq:M0}) and (\ref{eq:Mt}) after performing integration.   The
result is:
\begin{leftsubeqnarray}
\slabel{eq:stg2}
&& s\left(\frac\tau{\rm y}\right)=\frac{(m/m_\odot)^{2+p}-(m_{\rm mf}/m_\odot)
^{2+p}}{(m_{\rm Mf}/m_\odot)^{2+p}-(m_{\rm mf}/m_\odot)^{2+p}}~~;\qquad p\ne-2
~~; \\
\slabel{eq:ste2}
&& s\left(\frac\tau{\rm y}\right)=\frac{\ln(m/m_{\rm mf})}
{\ln(m_{\rm Mf}/m_{\rm mf})}~~;\qquad p=-2~~;
\label{seq:st}
\end{leftsubeqnarray}
regardless of the star formation efficiency.

\begin{figure}[t]  
\begin{center}      
\includegraphics[scale=0.8]{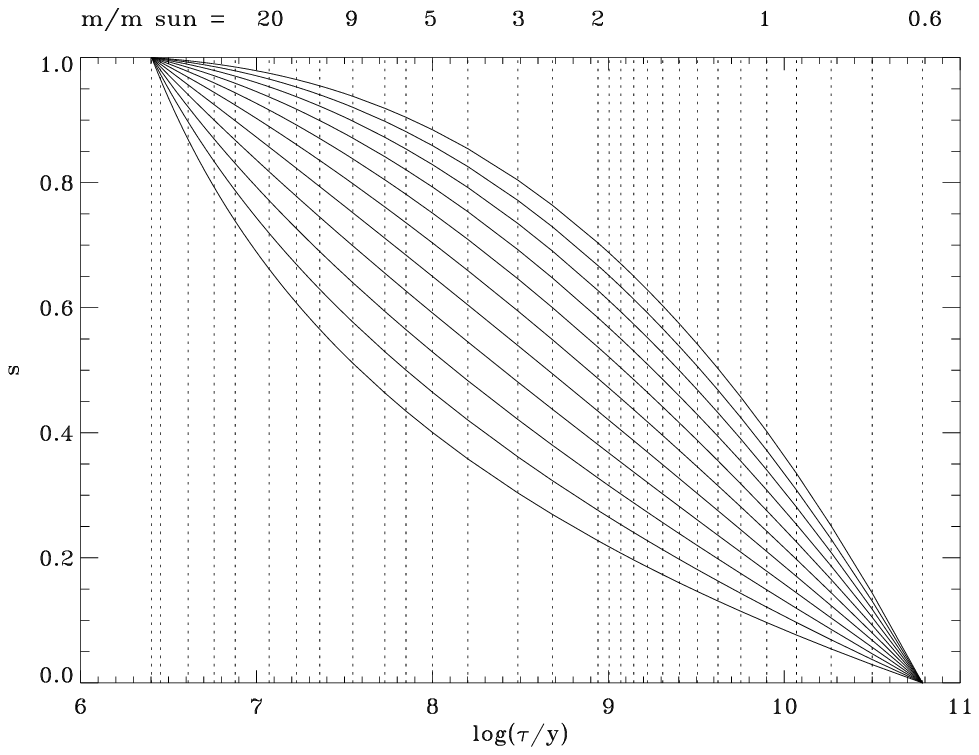}                      
\caption[ddbb]{Fractional mass of a star generation,
$s(\tau/{\rm y})=M^\ast(\tau/{\rm y})/M^\ast(0)$, as a function of the
logarithmic stellar lifetime, $\log(\tau/{\rm y})$, for a power-law stellar
initial mass function with exponent equal to (from top to bottom)
$p=-3.0, -2.9, ..., -2.0,$ with regard to a computed
$\log(\tau/{\rm y})$-$\log(m/m_\odot)$ relation interpolated by a simple
exponential,
restricted to the domain, $0.6\le m/m_\odot\le120.0$.   Related stellar
lifetimes, corresponding to the whole range for which stellar evolution has
been computed (Portinari et al. 1998), are marked by dotted vertical lines, a
restricted number of which are captioned on the top of the box.}
\label{f:fras}     
\end{center}       
\end{figure}

The fractional mass of a star generation, expressed by Eq.\,(\ref{seq:st}), is
plotted in Fig.\,\ref{f:fras} for $p=-3.0, -2.9, ..., -2.0,$ from top to
bottom, with regard to the interpolation curve expressed by a simple
exponential, Eq.\,(\ref{eq:yexp}), case 7 of Table \ref{t:pain}. 
Related stellar
lifetimes, corresponding to the whole range for which stellar evolution has
been computed in the parent paper (Portinari et al. 1998), are marked by
dotted vertical lines where $m_{\rm mf}/m_\odot=0.6$ and
$m_{\rm Mf}/m_\odot=120.0$.   A restricted number of stellar initial masses
are captioned on the top of the box.

An inspection of Fig.\,\ref{f:fras} shows the expected trend: mild stellar
initial mass functions $(p\appleq-2)$ imply a larger fraction of short-lived
stars and, in turn, a change in star mass fraction at a decreasing rate.   On
the
other hand, steep stellar initial mass functions $(p\appgeq-3)$ imply a larger
fraction of long-lived stars and, in turn, a change in star mass fraction at
an increasing rate.   Interestingly, the transition case where the
$s(\tau/{\rm y})$-$\log(\tau/{\rm y})$ relation is linear, is close to the
Salpeter's value, $p=p_{\rm lin}\approx-2.35$.

To gain more insight, the above mentioned special case relates to a straight
line passing through the points, $[\log(\tau_{\rm mf}/{\rm y}),0]$ and
$[\log(\tau_{\rm Mf}/{\rm y}),1]$, which can be expressed as:
\begin{equation}
\label{eq:stln}
s\left(\frac\tau{\rm y}\right)=\frac
{\log(\tau/{\rm y})-\log(\tau_{\rm mf}/{\rm y})}
{\log(\tau_{\rm Mf}/{\rm y})-\log(\tau_{\rm mf}/{\rm y})}~~;
\end{equation}
where $\tau_{\rm mf}=\tau(m_{\rm mf})$, $\tau_{\rm Mf}=\tau(m_{\rm Mf})$, for
brevity.   The combination of Eqs.\,(\ref{eq:stg2}) and (\ref{eq:stln}), after
little algebra, yields:
\begin{lefteqnarray}
\label{eq:taum2}
&& \log\left(\frac\tau{{\rm y}}\right)=\frac{\log(\tau_{\rm Mf}/{\rm y})-
\log(\tau_{\rm mf}/{\rm y})}{(m_{\rm Mf}/m_\odot)^{2+p}-
(m_{\rm mf}/m_\odot)^{2+p}}\left(\frac m{m_\odot}\right)^{2+p} \nonumber \\
&& \phantom{\log\left(\frac\tau{{\rm y}}\right)=}
-\frac{\log(\tau_{\rm Mf}/{\rm y})-
\log(\tau_{\rm mf}/{\rm y})}{(m_{\rm Mf}/m_\odot)^{2+p}-
(m_{\rm mf}/m_\odot)^{2+p}}\left(\frac{m_{\rm mf}}{m_\odot}\right)^{2+p}+
\log\left(\frac{\tau_{\rm mf}}{\rm y}\right)~~;\qquad
\end{lefteqnarray}
where $p=p_{\rm lin}$ via Eq.\,(\ref{eq:stln}).

The comparison of Eq.\,(\ref{eq:taum2}) with (\ref{eq:logtau}) term by term
implies the validity of the following relations:
\begin{lefteqnarray}
\label{eq:Cp2p}
&& C_2^\prime=-2-p~~; \\
\label{eq:Cp1p}
&& C_1^\prime=\frac{\log(\tau_{\rm Mf}/{\rm y})-
\log(\tau_{\rm mf}/{\rm y})}{(m_{\rm Mf}/m_\odot)^{2+p}-
(m_{\rm mf}/m_\odot)^{2+p}}~~; \\
\label{eq:C3p}
&& C_3=-\frac{\log(\tau_{\rm Mf}/{\rm y})-
\log(\tau_{\rm mf}/{\rm y})}{(m_{\rm Mf}/m_\odot)^{2+p}-
(m_{\rm mf}/m_\odot)^{2+p}}\left(\frac{m_{\rm mf}}{m_\odot}\right)^{2+p}+
\log\left(\frac{\tau_{\rm mf}}{\rm y}\right)~~;\qquad
\end{lefteqnarray}
where, in the case under discussion, the left-hand side of
Eqs.\,(\ref{eq:Cp2p})-(\ref{eq:C3p}) is known via Eq.\,(\ref{eq:Cpv}) and
$m_{\rm mf}=0.6$, $m_{\rm Mf}=120.0$.    Then $p=p_{\rm lin}$ can be evaluated
via Eqs.\,(\ref{eq:Cpv}), (\ref{eq:Cp2p}), and Eqs.\,(\ref{eq:Cp1p}) and
(\ref{eq:C3p}) are merely to be verified.   The result
is:
\begin{equation}
\label{eq:plin}
p_{\rm lin}=-2.374616~~;
\end{equation}
where the values of $C_1^\prime$ and $C_3$ coincide with their counterparts
expressed by Eq.\,(\ref{eq:Cpv}), as expected.

In conclusion, a star generation where the initial mass function is expressed
by a power-law within the range, $0.6\le m/m_\odot\le120.0$, implies a linear
dependence of the star mass fraction, $s(\tau/{\rm y})$, on the logarithmic
stellar lifetime, $\log(\tau/{\rm y})$, for an exponent of the initial mass
function close to Salpeter's value, $p_{\rm lin}\approx-2.35$.

\section{Discussion and conclusion} \label{dico}

Computed stellar $\log(\tau/{\rm y})$-$\log(m/m_\odot)$ relations are
interpolated to a good extent by double power-laws and, in particular, by a
single power-law, Eqs.\,(\ref{eq:y}) and (\ref{eq:yexp}), respectively,
regardless of the initial stellar metallicity, within the stellar initial mass
range, $0.6\le m/m_\odot\le120.0$.   The case of a single power-law is
expecially attractive due to its simplicity over an extended mass domain.

By comparison, a $\log(\tau/{\rm y})$-$\log(m/m_\odot)$ relation can be used
for massive $(m/m_\odot>6.6)$ stars (e.g., Matteucci and Greggio 1986), while
a cumbersome expression is necessary for less massive stars (e.g., Renzini and
Buzzoni 1986).   In alternative, four different
$\log(\tau/{\rm y})$-$\log(m/m_\odot)$ relations can be used within the mass
range, $1.3\le m/m_\odot\le120.0$ (Maeder and Meynet 1989).   A
three-parameter (depending on the stellar initial metallicity) fit to earlier
results from the Padua group (Alongi et al. 1993; Bressan et al. 1993;
Bertelli et al. 1994) is restricted to $Z\le0.03$ and the extrapolation does
not reproduce the correct trend as $m/m_\odot\to+\infty$ (Raiteri et al.
1996).

With regard to the interpolation curve, expressed by Eq.\,(\ref{eq:tame}), the
following relations hold for stellar initial masses outside the domain of
computed stellar evolution (Portinari et al. 1998) used in the current
attempt:
\begin{lefteqnarray}
\label{eq:tau0}
&& \log\left(\frac{\tau_0}{\rm y}\right)\to+\infty~~;\qquad\frac m{m_\odot}\to
0~~; \\
\label{eq:taui}
&& \log\left(\frac{\tau_\infty}{\rm y}\right)=C_3=5.70612~~;\qquad\frac m
{m_\odot}\to+\infty~~;
\end{lefteqnarray}
which is the correct trend, in the sense that (failed) stars with mass,
$m/m_\odot<m_{\rm mf}/m_\odot\approx0.08$, never start hydrogen burning and
related lifetime can be conceived as infinite.   On the other hand, stars
with increasingly high mass exhibit finite lifetime regardless of initial
metallicity.   The comparison between computed lifetimes (Portinari et al.
1998) and their counterparts inferred from the interpolation curve, expressed
by Eqs.\,(\ref{eq:yexp}), (\ref{eq:Cpv}), case 7 of Table \ref{t:pain}, is
shown in Table \ref{t:tmh}.
\begin{table}
\caption{Comparison between computed stellar lifetimes
(Portinari et al. 1998), $\tau_{\rm Pa98}$, and their counterparts inferred
from the single power-law interpolation curve, case 7 of Table \ref{t:pain},
$\tau_7$, outside the domain towards high-mass stars.   See text for further
details.}
\label{t:tmh}
\begin{center}
\begin{tabular}{rll} \hline
\multicolumn{1}{c}{$m/m_\odot$} &
\multicolumn{1}{c}{$\tau_{\rm Pa98}/{\rm y}$} &
\multicolumn{1}{c}{$\tau_7/{\rm y}$} \\
\hline
 150      & 3.0E6 & 2.2286E6 \\
 200      & 2.7E6 & 1.9162E6 \\
 300      & 2.5E6 & 1.5894E6 \\
 500      & 2.0E6 & 1.3032E6 \\
1000      & 2.0E6 & 1.0507E6 \\
$+\infty$ &       & 5.0830E5 \\
\hline
\end{tabular}                     
\end{center}                      
\end{table}                       
It is apparent lifetimes calculated via the single power-law interpolation
curve outside the domain towards high stellar masses are understimated with
respect to computed values, by a factor not exceeding 2 up to
$m/m_\odot=1000$.

Concerning stars with decreasingly low mass, a similar comparison is shown in
Table \ref{t:tml}, where computed stellar lifetimes are from a different
source (Laughlin et al. 1997), restricted to stellar initial metallicity,
$Z=0.02$ (Laughlin and Bodenheimer 1993).
\begin{table}
\caption{Comparison between computed stellar lifetimes
(Laughlin et al. 1997), $\tau_{\rm La97}$, and their counterparts inferred
from the single power-law interpolation curve, case 7 of Table \ref{t:pain},
$\tau_7$, outside the domain towards low-mass stars.   See text for further
details.}
\label{t:tml}
\begin{center}
\begin{tabular}{rll} \hline
\multicolumn{1}{c}{$m/m_\odot$} &
\multicolumn{1}{c}{$\tau_{\rm La97}/{\rm y}$} &
\multicolumn{1}{c}{$\tau_7/{\rm y}$} \\
\hline
0.08 & 1.1E13 & 3.24E16 \\
0.10 & 6.2E12 & 4.40E15 \\
0.12 & 4.2E12 & 9.72E14 \\
0.14 & 3.4E12 & 2.93E14 \\
0.16 & 2.7E12 & 1.10E14 \\
0.18 & 2.0E12 & 4.78E13 \\
0.20 & 1.8E12 & 2.35E13 \\
0.25 & 1.7E12 & 5.71E12 \\
\hline
\end{tabular}                     
\end{center}                      
\end{table}                       
It is apparent lifetimes calculated via the single power-law interpolation
curve outside the domain towards low stellar masses are unacceptably
overstimated with respect to computed values but, interestingly, the
discrepancy is within the same order of magnitude for $m/m_\odot=0.25$, where
stellar evolution still attains the giant phase (Laughlin et al. 1997).
Accordingly, an absence of the giant phase would imply a different
$\log(\tau/{\rm y})$-$\log(m/m_\odot)$ relation.

A simple but acceptable linear fit to the results listed in Table \ref{t:tml},
leaving aside $m/m_\odot=0.25$, reads:
\begin{equation}
\label{eq:linf}
\log\left(\frac\tau{\rm y}\right)=-2\log\left(\frac m{m_\odot}\right)+10.8~~;
\end{equation}
which implies $\log(\tau_{1.0}/{\rm y})=10.800$ and 
$\log(\tau_{0.6}/{\rm y})=11.257$, larger but of about the same order of
magnitude with respect to their counterparts determined from stellar
evolution (Portinari et al. 1998).

\begin{figure}[t]  
\begin{center}      
\includegraphics[scale=0.8]{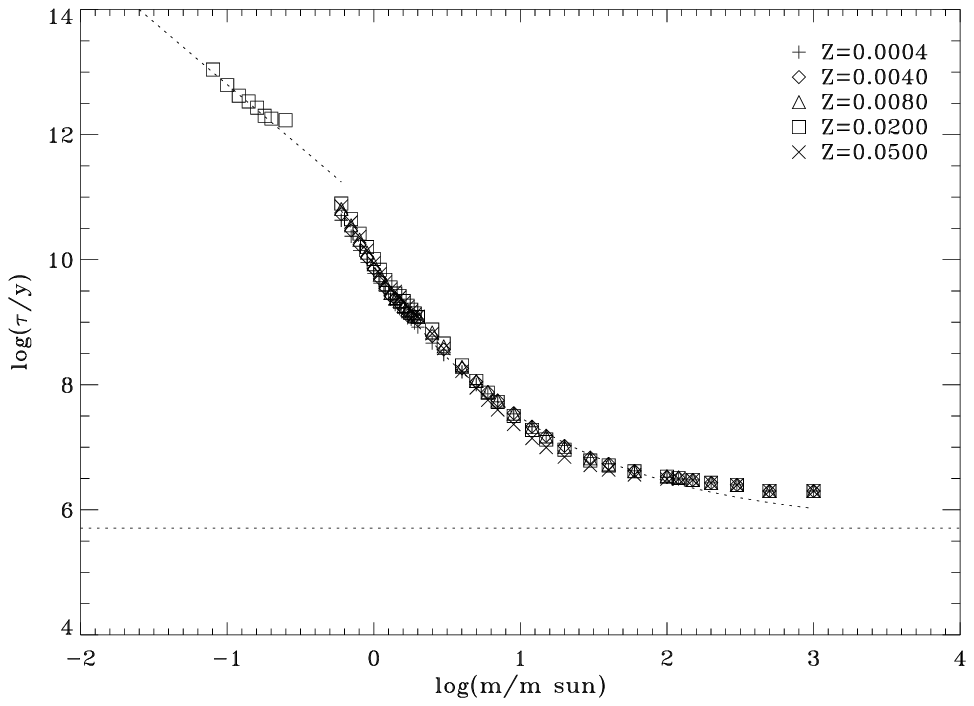}                      
\caption[ddbb]{Computed $\log(\tau/{\rm y})$-$\log(m/m_\odot)$ relation,
plotted
in Fig.\,\ref{f:stelle}, extended to both lower (Laughlin et al. 1997) and
larger (Portinari et al. 1998) stellar initial masses.   Lower masses relate
to stellar initial metallicities, $Z=0.02$.   An interpolation straight line
and power-law curve, respectively, are also shown (dotted).   The asymptotic
limit of the interpolation curve towards infinite stellar initial mass is
marked by a dotted horizontal line.   See text for further details.}
\label{f:stellp}     
\end{center}       
\end{figure}

The computed $\log(\tau/{\rm y})$-$\log(m/m_\odot)$ relation, represented in
Fig.\,\ref{f:stelle}, extended to the results listed in Tables \ref{t:tmh}
and \ref{t:tml}, is shown in Fig.\,\ref{f:stellp} where, in addition, the
single power-law interpolation curve and the interpolation line, 
Eqs.\,(\ref{eq:yexp}), (\ref{eq:Cpv}), and (\ref{eq:linf}), respectively, are
also plotted together with the asymptotic limit of the interpolation curve for
infinite stellar initial mass.
The global trend exhibited by the computed
$\log(\tau/{\rm y})$-$\log(m/m_\odot)$ relation is reminiscent of earlier
results restricted to stellar initial metallicity, $Z=0.02$ (Maeder and Meynet
1989), where the transition between two different regimes occurs above (e.g.,
Wiersma et al. 2009) instead of below (Fig.\,\ref{f:stellp}) the solar mass.

In conclusion, the single power-law interpolation curve, expressed by
Eqs.\,(\ref{eq:yexp}), (\ref{eq:Cpv}), can safely be extrapolated towards
high-mass stars $(m/m_\odot>120.0)$ yielding understimated values within a
factor of two up to $m/m_\odot=1000$ and within a fiducial factor of four up
to $m/m_\odot\to+\infty$.   Additional caution must be used for the
extrapolation towards low-mass stars $(m/m_\odot<0.6)$, yielding overstimated
values within a factor of about three down to $m/m_\odot=0.25$ and
to unacceptably larger overstimates down to $m/m_\odot=0.08$.

The question if the single power-law interpolation curve, expressed by
Eqs.\,(\ref{eq:yexp}), (\ref{eq:Cpv}), arises from a mere coincidence or has
a theoretical interpretation, is outside the aim of the current attempt but a
few considerations can be performed.   Differentiating both sides of 
Eq.\,(\ref{eq:yexp}), after little algebra yields:
\begin{lefteqnarray}
\label{eq:tmd}
&& \frac{\diff\stau}{\diff\semme}=-\lambda\semme^{-\beta}\stau~~; \\
\label{eq:pmd}
&& \stau=\frac\tau{\rm y}~~;\qquad\semme=\frac{m}{m_\odot}\qquad
\lambda=C_1^\prime C_2^\prime~~;\qquad\beta=C_2^\prime+1~~;
\end{lefteqnarray}
accordingly,  longer stellar lifetimes imply larger change of stellar lifetime with stellar initial mass
and vice versa.   The special case, $\beta=0$, or $C_2^\prime=-1$, reduces to
the equation of the exponential decay, $\stau$ being related to the number of
decaying nuclides and $\semme$ to the time.   On the other hand,
$C_2^\prime=-1$ would imply stellar lifetime proportional to initial stellar
mass via Eq.\,(\ref{eq:logtau}), which is ruled out by the theory of stellar
evolution.
         
The fractional stellar mass as a function of the logarithmic stellar lifetime,
shown in Fig.\,\ref{f:fras}, relates to a single star generation and to a
power-law stellar initial mass function, which are limiting cases.   In
reality, multiple star generations appear even in low-mass stellar systems
and the stellar initial mass function exhibits a more complex trend with
respect to a simple power-law.   On the other hand, the ideal situation
depicted in Fig.\,\ref{f:fras} is expected to show a similar trend with
respect to the real case, at least for environments where the whole amount of
pristine gas was turned into stars in a short time, less than 1 Gyr say, such
as the inner halo, the thick disk, and globular clusters.   It is worth
noticing the stellar mass is less than the total mass, in that the
contribution of stellar remnants (white dwarfs, neutron stars, black holes)
and baryonic dark matter is neglected: in any case, the related contribution
is expected to be of a few percent at most.

With regard to globular clusters with age equal to $\tau_{0.9}$ (first
vertical line on the right of $\log(\tau/{\rm y})=10$), an inspection of
Fig.\,\ref{f:fras} shows a present star mass fraction, $0.07\appleq s\appleq
0.35$, according if $-2\ge p\ge-3$, in particular $s\approx0.15$ for the
Salpeter's exponent, $p=-2.35$, close to a linear dependence of the star mass
fraction on the logarithmic stellar lifetime.   Then a mild stellar initial
mass function, $p\approx-2.1$, implies an initial mass larger by a factor of
about ten, for globular clusters with age equal to $\tau_{0.9}$, in the case
under consideration.

The main results of the current note can be summarized as follows.
\begin{description}
\item[(1)\hspace{2.0mm}]
The computed $\log(\tau/{\rm y})$-$\log(m/m_\odot)$ relation for the stellar
initial mass range, $0.6\le m/m_\odot\le120.0$, and the stellar initial
metallicity range, $0.0004\le Z\le0.0500$, tabulated in an earlier attempt
(Portinari et al. 1998) is fitted to a good extent by a double power-law for
assigned initial metallicity, which can be reduced to a single power-law for
the whole set of initial metallicities.   The relative errors,
$R[\log(\tau/{\rm y})]=1-\log(\tau_{\rm fit}/{\rm y})/\log(\tau/{\rm y})$, do
not exceed about 2\% for the double power-law and 4\% for the single
power-law.
\item[(2)\hspace{2.0mm}]
The interpolation curve, expressed by a single power-law, can be extrapolated
towards both high-mass and low-mass stars.    In the former alternative, 
stellar lifetimes are understimated by a factor less than 2 up to $m/m_\odot=
1000$ (Portinari et al. 1998) and by a fiducial factor less than 4 up to
$m/m_\odot\to+\infty$.   In the latter alternative, stellar lifetimes are
overstimated by a factor less than about 3 down to $m/m_\odot=0.25$ and by an
unacceptably large factor down to $m/m_\odot=0.08$ (Laughlin et al. 1997).
\item[(3)\hspace{2.0mm}]
In the special case of a single star generation with stellar initial mass
function defined by a power-law, using the single power-law interpolation
curve within the mass range, $0.6\le m/m_\odot\le120.0$, the star mass
fraction declines in time at a decreasing rate for mild stellar initial mass
function $(p\appleq-2)$ and at an increasing rate for steep stellar initial
mass function $(p\appgeq-3)$, where a linear trend is exhibited for a value
of the exponent close to the Salpeter's value $(p_{\rm lin}\approx-2.35)$.
\end{description}

\section*{Acknowledgement}
Thanks are due to C. Chiosi for useful discussions.


\begin{thebibliography}{}
%

\bibitem{} Acharova, I.A., Gibson, B.K., Mishurov, Yu.N., Kovtyukh, V.V.:
           2013, Astron. Astrophys., 557, A107.


\bibitem{} Alongi, M., Bertelli, G., Bressan, A., et al.:
           1993, Astron. Astrophys. Supp, 97, 851.


\bibitem{} Asplund, M., Grevesse, N., Sauval, A.J., Scott, P.: 2009,
                 Ann. Rew. Astron. Astrophys., 47, 481.


\bibitem{} Bertelli, G., Bressan, A., Chiosi, C., Fagotto, F., Nasi, E.: 1994,
           Astron. Astrophys. Supp., 106, 275.


\bibitem{} Bressan, A., Fagotto, F., Bertelli, G., Chiosi, C.:
           1993, Astron. Astrophys. Supp, 100, 647.


\bibitem{} Caimmi, R.: 2006, App. Math. and Comp., 174, 447.


\bibitem{} Caimmi, R.: 2011, Ser. Astron. J., 183, 37.




\bibitem{} Laughlin, G., Bodenheimer, P: 1993, Astrophys. J., 403, 303.

\bibitem{} Laughlin, G., Bodenheimer, P, Adams, F.C.: 1997, Astrophys. J.,
           482, 420.


\bibitem{} Maeder A., Meynet G.: 1989, Astron. Astrophys., 210, 155.


\bibitem{} Matteucci, F., Greggio, L.: 1986, Astron. Astrophys., 154, 279.


\bibitem{} Padovani P., Matteucci M.: 1993, Astrophys. J., 416, 26.



\bibitem{} Pagel, B.E.J., Patchett, B.E.: 1975, Mon. Not. R. Astron. Soc.,
                172, 13.


\bibitem{} Portinari, L., Chiosi, C., Bressan, A.: 1998, Astron. Astrophys.,
           334, 505.


\bibitem{} Raiteri, C.M., Villata, M., Navarro, J.E.: 1996, Astron.
           Astrophys., 315, 105.


\bibitem{} Renzini, A., Buzzoni, A.: 1986, in Spectral Evolution of Galaxies,
                 ed. C. Chiosi and A. Renzini (Dordrecht: Reidel), 195.


\bibitem{} Romano D., Chiappini C., Matteucci F., Tosi M.: 2005,
           Astron. Astrophys., 430, 491.


\bibitem{} Salpeter, E.E.: 1955, Astrophys. J., 121, 161.


\bibitem{} Wiersma, R.P.C., Schaye, J., Theuns, T., Dalla Vecchia, C.,
           Tornatore, L.: 2009, Mon. Not. R. Astron. Soc., 399, 574.

\end{thebibliography}
\end{document}